\documentclass[fleqn,10pt]{wlscirep}

\usepackage[square, comma, sort&compress, numbers]{natbib}

\usepackage{amsmath}
\usepackage{amssymb}
\usepackage{upgreek}
\usepackage{url}
\usepackage[T1]{fontenc}
\usepackage{ae,aecompl}
\usepackage{hyperref}
\usepackage{adjustbox}
\usepackage[normalem]{ulem}

\title{HD\,66051, an eclipsing binary hosting a highly peculiar, HgMn-related star} 

\author[1,*]{Ewa Niemczura}
\author[2,3]{Stefan H{\"u}mmerich}
\author[4]{Fiorella Castelli}
\author[5]{Ernst Paunzen}
\author[2,3]{Klaus Bernhard}
\author[2,3,6]{Franz-Josef Hambsch}
\author[7]{Krzysztof He\l{}miniak}

\affil[1]{Instytut Astronomiczny, Uniwersytet Wroc{\l}awski, PL-51-622 Wroc{\l}aw, Poland}
\affil[2]{American Association of Variable Star Observers (AAVSO), 49 Bay State Rd, Cambridge, MA 02138, USA}
\affil[3]{Bundesdeutsche Arbeitsgemeinschaft f{\"u}r Ver{\"a}nderliche Sterne e.V. (BAV), D-12169 Berlin, Germany}
\affil[4]{Instituto Nazionale di Astrofisica, Osservatorio Astronomico di Trieste, Via Tiepolo 11, I-34143 Trieste, Italy}
\affil[5]{Department of Theoretical Physics and Astrophysics, Masaryk University, Kotl\'a\v{r}sk\'a 2, CZ-611 37 Brno, Czech Republic}
\affil[6]{Vereniging Voor Sterrenkunde (VVS), Brugge, BE-8000, Belgium}
\affil[7]{Department of Astrophysics, Nicolaus Copernicus Astronomical Center, ul. Rabia\'nska 8, PL-87-100 Toru\'n, Poland}

\affil[*]{niemczura@astro.uni.wroc.pl}

\begin{abstract} 
HD\,66051 is an eclipsing system with an orbital period of about 4.75\,d that exhibits out-of-eclipse variability with the same period. New multicolour photometric observations confirm the longevity of the secondary variations, which we interpret as a signature of surface inhomogeneities on one of the components. Using archival and newly acquired high-resolution spectra, we have performed a detailed abundance analysis. The primary component is a slowly rotating late B-type star (\teff\,=\,$12500\pm200$\,K; \logg\,=\,$4.0$, \vsini\,=\,$27\pm2$\,\kms) with a highly peculiar composition reminiscent of the singular HgMn-related star HD\,65949, which seems to be its closest analogue. Some light elements as He, C, Mg, Al are depleted, while Si and P are enhanced. Except for Ni, all the iron-group elements, as well as most of the heavy elements, and in particular the REE elements, are overabundant. The secondary component was estimated to be a slowly rotating A-type star (\teff\,$\sim8000$\,K; \logg\,=\,$4.0$, \vsini\,$\sim18$\,\kms). The unique configuration of HD\,66051 opens up intriguing possibilities for future research, which might eventually and significantly contribute to the understanding of such diverse phenomena as atmospheric structure, mass transfer, magnetic fields, photometric variability and the origin of chemical anomalies observed in HgMn stars and related objects.

\end{abstract}

\begin{document}

\def\teff{${T}_{\rm eff}$}
\def\kms{{km\,s}$^{-1}$}
\def\logg{$\log g$}
\def\micro{$\xi_{\rm t}$}
\def\vsini{$v\sin i$}
\def\ebv{$E(B-V)$}
\newcommand{\ion}[2]{{#1}\,{\sc #2}}

\flushbottom
\maketitle

\thispagestyle{empty}

\section*{Introduction} 

Chemically peculiar (CP) stars comprise about 10\,\% of upper main-sequence stars (spectral types early B to early F). They are characterized by peculiar atmospheric abundances which deviate significantly from the solar composition. In some CP stars, excesses up to several orders of magnitude are found among elements like Si, Hg, or the rare-earth elements (REE), while the atmospheres of other groups are characterized by the depletion of certain elements, as has been observed e.g. in the He-weak stars \citep{preston74}. For most groups of CP stars, the observed chemical peculiarities are attributed to atomic diffusion, i.e. the interplay of radiative levitation and gravitational settling operating in the calm radiative atmospheres of slowly rotating stars \citep{michaud70,richer00}. While most elements sink under the influence of the gravitational force, other elements, which have absorption lines near the local flux maximum, are radiatively driven outward.

Two groups of CP stars are relevant to the present investigation. The Bp/Ap stars are set apart by the presence of stable, globally-organized magnetic fields with strengths from about 300\,G to several tens of kG \citep{babcock47,auriere07}. The origin of the observed magnetic fields is still a matter of some controversy \citep{moss04}, the main competing theories being the dynamo theory (field generated by dynamo action in the convective core) and the fossil field theory (field is a relic of the 'frozen-in' interstellar magnetic field). Although evidence strongly favours the fossil field theory \citep{braithwaite04}, dynamo action in some mass regimes has been postulated \citep{hubrig07}. For convenience, these objects are referred to hereafter as Ap stars.

Due to the influence of the magnetic field, Ap stars exhibit a non-uniform distribution of chemical elements on their surfaces \citep{michaud81}. Flux is redistributed in these abundance spots and patches as a result of bound-free and bound-bound atomic transitions \citep{molnar73,lanz96,krticka13}. As has been described in the oblique rotator model \citep{stibbs50}, the magnetic axis is not aligned with the rotational axis. Therefore, as a result of rotation, strictly periodic changes are observed in the spectra and brightness of many Ap stars. The spot configurations on Ap stars remain stable for decades and more, which is probably a consequence of the magnetic field.

Another important group is made up of the so-called Mercury-Manganese (HgMn) stars \citep{preston74}. As their name implies, these stars are characterized by their unusually strong lines of Hg and Mn. Numerous other peculiarities -- like He, Ni or Al deficiency and overabundances of elements such as P, Ga and Xe -- are regularly observed \citep{castelli04}. HgMn stars generally show increasing overabundance of heavy elements with atomic number and are encountered in the temperature range of 10\,000\,K\,$\leq$\,\teff\,$\leq\,$16\,000\,K \citep{smith96, ghazaryan16}. The observed abundance ratios may differ significantly from one star to another, and it is not understood why the diffusion processes thought to operate in these stars create different abundance patterns in objects of similar temperature \citep{cowley14}. HgMn stars are often found in binary and multiple systems. In fact, a binarity fraction of more than 50\,\% among this group of stars has been established \citep{smith96}, with some studies indicating rates as high as 91\,\% \citep{schoeller10}. Binarity has been proposed to play a vital role in the development and understanding of the observed abundance patterns in HgMn stars.

HgMn stars do not show strong, organized magnetic fields, but the presence of weak fields has been claimed by several investigators \citep{hubrig12}. This, however, remains controversial. It has been shown that HgMn stars show an inhomogeneous surface distribution of elements exhibiting signs of secular evolution \citep{kochukhov07}. Only very few HgMn stars are known to be photometric variables.

The peculiarities observed in CP stars range from very mild to extreme \citep{loden87}. Among the most peculiar objects, there are the extreme lanthanide stars HD\,51418 \citep{jones74}, HD\,9996 \citep{preston70} and the HgMn-related object HD\,65949 \citep{cowley10}. Of special notice is HD\,101065 \citep[Przybylski's star,][]{przybylski66}, which is widely regarded as the most peculiar star known.

CP stars have a long history of study and are important to stellar astrophysics in several respects. As indicated above, these objects continue to baffle theoreticians and defy easy abundance analyses and classification. Their complex atmospheres offer generally applicable insight on (the interplay of) such diverse phenomena as magnetic fields, atomic diffusion, stellar rotation, pulsational variability \citep[observed among the so-called rapidly oscillating Ap (roAp) stars, see][]{kurtz82}, vertical abundance stratification and non-standard temperature gradients. Therefore, by studying and deciphering the atmospheres of these peculiar objects, valuable information can be gained for the general understanding and modelling of stars.

The present work presents a detailed analysis of HD\,66051, which has been found to be an eclipsing binary system containing a highly peculiar CP star.

\section*{Results}

\section*{An eclipsing binary system}

The photometric variability of HD\,66051 (V414\,Pup; RA,\,Dec\,(J2000)\,=\,08h\,01m\,24s.642,\,-12$^\circ$\,47'\,35''.76) was discovered in Hipparcos data \citep{vanleeuwen97}. However, because of an insufficient number of observations, no variability type could be determined. The star was subsequently discovered to be an eclipsing binary of Algol-type with an orbital period of $P$\,=\,4.74922\,d and a magnitude range of 8.79--9.12\,mag ($V$) using Hipparcos and ASAS-3 data \citep{otero03,huemmerich16}. Additional, out-of-eclipse variability with the same period was established and interpreted as being due to rotationally-induced variability caused by surface inhomogeneities on one of the system's synchronously rotating components.

\begin{figure}[ht]
\includegraphics[width=0.5\linewidth]{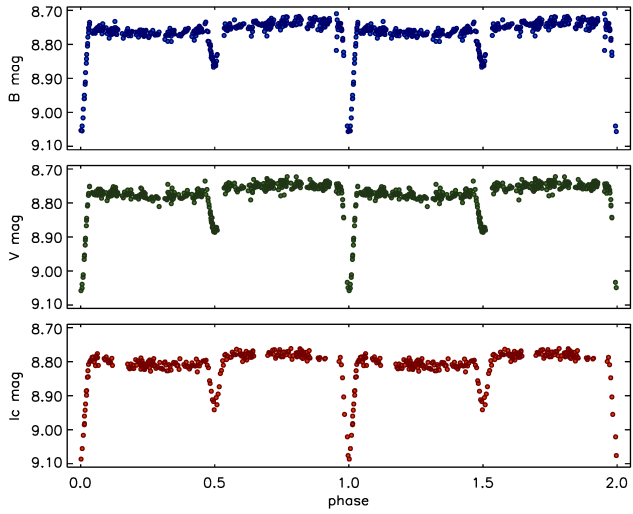}
\caption{Phase plots of HD\,66051, based on our own photometric observations. The panels indicate, from top to bottom, $B$, $V$, and $I_C$ data, respectively. Data have been binned (bin-size: $0.02$\,d). The secondary variability is clearly visible in all datasets.}
\label{pp_ROAD}
\end{figure}

We have acquired new extensive multicolour photometric observations that confirm the results from the sky survey data and allow a more detailed study. A total number of $2776$, $2803$, and $2271$  observations were acquired during a timespan of 60, 60, and 37 days in, respectively, $V$, $B$, and $I_C$ (see Methods: Photometric data -- acquisition and analysis). The photometric data are available as supplementary material to this paper. The newly acquired data corroborate the findings of the aforementioned investigators. Furthermore, from an analysis of all available data and the significantly improved time-span of the observations, a refined ephemeris has been derived (${\rm\,Min(HJD)}=2452167.867(2)+4.749218(2)\times{\rm\,E}$; phase zero corresponds to the time of the primary eclipse). Phase plots based on these data are presented in Fig.\,\ref{pp_ROAD}.

The recent observations confirm the longevity of the observed secondary variability in the light curve, which remained stable from the beginning of Hipparcos coverage in 1990 to the present date. This may be explained by synchronous rotation of the variable component in the system, i.e. both stars are tidally locked and the variability always occurs at exactly the same phase. The observed rotational period agrees with the projected rotational velocity of $27\pm2$\,\kms\ deduced from the spectroscopic analysis.

\section*{Spectroscopic analysis} 

HD\,66051 has been previously found to exhibit enhanced Si lines. As this is a characteristic of the magnetic chemically peculiar stars, it was consequently classified as an Ap star \citep[spectral type ApSi;][]{bidelman73,houk88}. The 'Catalogue of Ap, HgMn and Am stars' indicates a spectral type of A0pSi \citep{RM09}. No further detailed studies of HD\,66051 exist.

However, a high-resolution spectrum of the star is available in the archive of the 'Variable Star One-shot Project' \citep{dall07}, which was taken with the HARPS instrument \citep[High Accuracy Radial velocity Planet Searcher;][]{HARPS} at the ESO La Silla 3.6\,m telescope in Chile. The star was observed on JD\,2453827.518802, which corresponds to orbital phase $\varphi_{orb}$\,=\,0.46. Additionally, we observed the star with the HIDES instrument \citep[HIgh-Dispersion Echelle Spectrograph;][]{HIDES} at the 1.88\,m telescope of the Okayama Astrophysical Observatory (OAO) in Japan. This spectrum was obtained on JD\,2457817.06220, corresponding to $\varphi_{orb}$\,=\,0.50. The HIDES spectrum is available as supplementary material to this paper.

Therefore, the HIDES spectrum was taken during secondary eclipse and is dominated by the light of the primary star. Nevertheless, the available material does not allow a detailed analysis of the system's parameters, and, at this stage, it is impossible to exclude a slight contribution from the secondary star to the spectrum. However, both components are considerably different in luminosity (see Section: Spectroscopic analysis), and this contribution, if present at all, will be weak at best. A solution of the system, based on additional spectroscopic material, will be presented in an upcoming paper.

The HIDES spectrum lacks numerous lines that are present in the composite HARPS spectrum and that we attribute to the secondary component (see Section: Secondary component). As the HARPS spectrum has higher signal-to-noise ratio and higher resolution than the HIDES spectrum, both spectra were used for spectral classification and analysis.

\subsection*{Spectral classification}

As an initial step, we have compared the spectrum to the standard star spectra recommended by \citep{gray09} and derived the spectral type of HD\,66051 from the hydrogen and \ion{Ca}{k} lines. Generally, for non-chemically peculiar stars, these sets of lines provide the same spectral type. For CP stars, however, different spectral types are commonly derived from both elements. The \ion{Ca}{k} line indicates a spectral type of B9. However, from an investigation of the hydrogen H$\beta$, H$\gamma$, and H$\delta$ lines, a spectral type of B7 has been derived. Furthermore, there is some ambiguity concerning the luminosity classification: while the H$\beta$ line indicates a main-sequence object of luminosity class V, the H$\gamma$ and H$\delta$ lines point to luminosity class III and therefore a more evolved object. This is the consequence of normalization problems of the broad and heavily blended H$\gamma$ and H$\delta$ wings typically encountered in CP stars, so we have rejected this result. From consideration of the \ion{Ca}{k} and hydrogen H$\beta$ line, we have thus derived a spectral type of kB9\,hB7\,V.

As a next step, the peculiarities of the spectrum were considered. HD\,66051 is clearly a He-weak star: the helium lines are very weak and barely seen in the spectrum. By contrast, it is immediately obvious that several elements are significantly overabundant. Enhanced lines of all elements usually used for spectral classification of CP stars, namely Si, Cr, Sr and Eu, were found (see Methods: Spectroscopic analysis). We also detected significantly enhanced lines of Hg but were unable to derive a possible Mn enhancement from the very rich spectrum. We thus arrived at a classification of kB9\,hB7\,V\,He-w\,SiCrSrEuHg. However, it is obvious that this classification only insufficiently describes the complex and peculiar spectrum of HD\,66051, which defies simple classification. In fact, it is well known that the simple peculiarity groups defined for classification in the past are oversimplifications of the real situation \citep{gray09} and there are stars so peculiar as to being unique and not fitting any of the customary groups. We therefore resorted to a more detailed abundance analysis, which is provided in the corresponding section below.

\subsection*{Atmospheric parameters}

Several independent methods were used to obtain atmospheric parameters for the primary star of HD\,66051. Effective temperature (\teff) and surface gravity (\logg) were calculated from calibrations of photometric indices and synthesis of hydrogen-line profiles. Because of the increased line blanketing in CP stars, specific corrections need to be taken into account in order to derive reliable values via the above mentioned techniques. An overview of the results is presented in Table\,\ref{atmospheric_parameters}.

\begin{figure}[ht]
\includegraphics[width=0.90\textwidth]{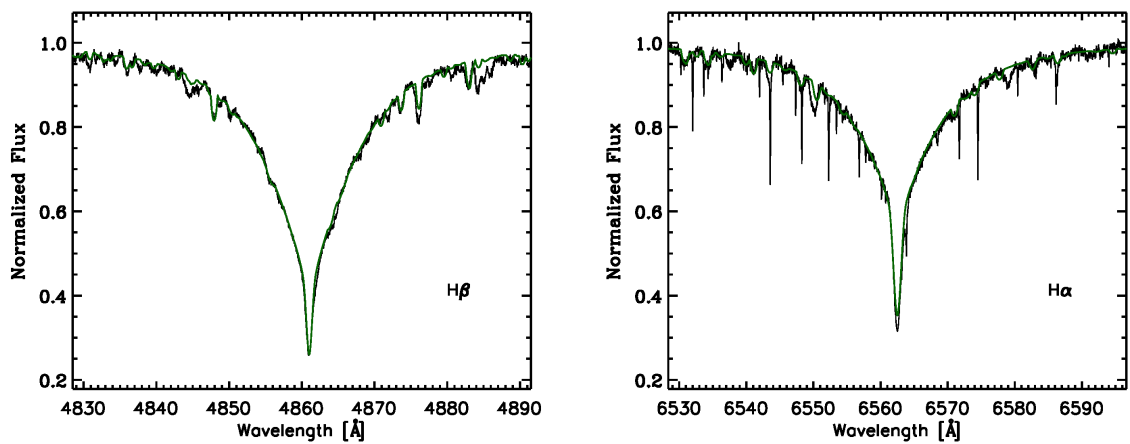}
\caption{Observed hydrogen H$\alpha$ and H$\beta$ lines (black lines) and best-fit synthetic profiles (green lines).}
\label{fig:balmers}
\end{figure}

The results derived from the photometric calibrations and the fitting of the hydrogen lines are in agreement with each other. For the abundance determination, we have chosen to adopt the atmospheric parameters derived from the hydrogen line fitting. In Fig.\,\ref{fig:balmers}, we present the observed Balmer lines and the synthetic profiles calculated for the derived atmospheric parameters. We verified that the effect of the secondary on the Balmer profiles is negligible. The parameters of the secondary star listed in the bottom row of Table\,\ref{atmospheric_parameters} are an estimate derived from the comparison of the observed and computed spectra (see Section: Secondary component).

\subsection*{Abundance analysis}

\begin{figure}[ht]
\includegraphics[width=0.90\textwidth]{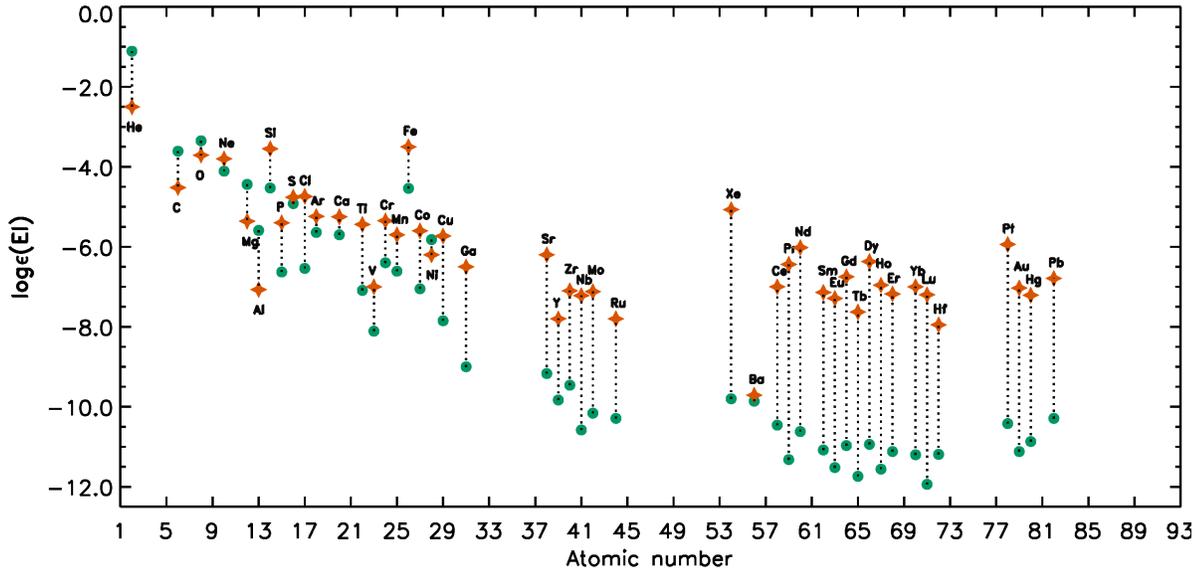}
\caption{Comparison of the chemical composition of HD\,66051 (orange stars) to the solar abundance pattern (green circles). The elements overabundant in HD\,66051 are signed above the symbol corresponding to the analysed star, whereas the underabundat elements are signed below the symbol.}
\label{fig:abundances2}
\end{figure}

The chemical peculiarities of HD\,66051 indicated by the spectral classification are fully confirmed by the abundance analysis. The derived abundances are presented in Table\,\ref{chemical_abundances}. A microturbulence of \micro\,=\,$0.5$\,\kms\ was assumed as typical of chemically peculiar stars \citep[see e.g.][]{catanzaro16}. The calculated projected rotational velocity equals $27\pm2$\,\kms. Fig.\,\ref{fig:abundances2} compares the derived chemical abundances with solar values adopted from \citep{asplund09}. While some elements are depleted relative to solar abundance, most elements are significantly enhanced. In fact, for atomic number $Z > 20$, all elements except Ni are overabundant; the strength of the enhancement generally increases with increasing $Z$.

\begin{figure}[ht]
\includegraphics[width=0.90\textwidth]{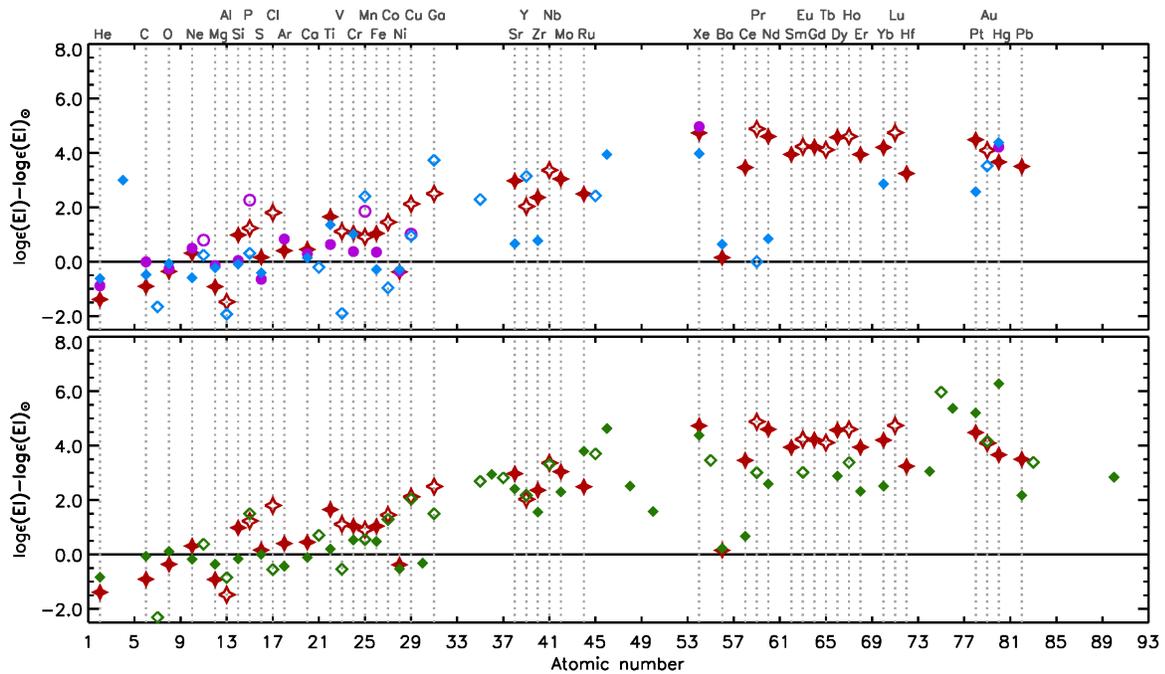}
\caption{Comparison of the chemical composition of HD\,66051 (red stars) to the abundance patterns observed in the HgMn stars HD\,49606 (pink circles), HD\,175640 (blue diamonds) [upper panel] and the HgMn-related, highly peculiar star HD\,65949 (green diamonds) [lower panel]. Odd-$Z$ elements and even-$Z$ elements are denoted by, respectively, open and filled symbols.}
\label{fig:abundances1}
\end{figure}

Interestingly, the detailed abundance analysis revealed similarities to the abundance pattern of the HgMn stars, which has never been reported in the literature. For the following discussion, we have therefore compared the derived chemical abundances of HD\,66051 with values of two thoroughly analysed HgMn stars: HD\,49606 \citep{catanzaro16} and HD\,175640 \citep{castelli04}, and the HgMn-related, highly peculiar star HD\,65949 \citep{cowley10}. The results are presented in Fig.\,\ref{fig:abundances1}. We have chosen these particular stars because they share similar atmospheric parameters with HD\,66051. \citep{castelli04} derived \teff\,=\,12000\,K and \logg\,=\,3.95\,dex for HD\,175640 from the $c_1$ and $\beta$ indices and assumed a microturbulence of \micro\,=\,$0.0$\,\kms. This star is rotating very slowly at \vsini\,$=2.5$\,\kms. The atmospheric parameters \teff\ and \logg\ of HD\,49606 were determined from Balmer line analysis as $13000\pm150$\,K and $3.80\pm0.05$\,dex, respectively. The value of \micro\ was obtained from Fe lines as $0.3^{+0.9}_{-0.0}$\,\kms. The projected rotational velocity of HD\,49606 equals $19.0\pm0.5$\,\kms\ \citep{catanzaro16}. \citep{cowley10} derived the atmospheric parameters of HD\,65949 from photometric indices and Balmer line analysis and adopted \teff\,=\,$13100$\,K and \logg\,=\,$4.0$\,dex for the abundance analysis.

In the following, similarities and differences to the group of the HgMn stars are discussed on the basis of the information provided by \citep{ghazaryan16}, who discuss the abundance anomalies in HgMn stars based on an exhaustive survey of the recent literature.

Helium is significantly underabundant in HD\,66051; its neutral lines are only very weakly present in the spectrum. The determined He abundance ($\log{\rm\,He/Ntot}=-2.50\pm0.10$) is only a rough estimate based on the fitting of $5$ \ion{He}{i} lines ($5875.6$, $4921.9$, $4713.1$, $4471.5$, and $4026.2$\,{\AA}), from which we deduce that He is approximately $25$ times less abundant than in the sun, as has been commonly observed in HgMn stars.

For the light elements, a differentiated abundance pattern has been found. Ca and O exhibit abundances close to solar values; N, however, is solar or  depleted, which is a general characteristic of the HgMn stars. Sulphur is underabundant in most HgMn stars, but is close to the solar value in HD\,66051, as in the case of HD\,65949. The most underabundant of these elements are Al and the $\alpha$ element Mg. While a strong deficiency of Al is typical of HgMn stars, Mg is usually less depleted. On the other hand, the strongest overabundances among this group of elements are observed for Cl, Si and P. The latter element is commonly enhanced in HgMn stars; Si, however, is not, which provides a contrast to the peculiarities observed in most HgMn stars. Interestingly, there is a discrepancy between the abundances derived from different Si lines, which may suggest vertical stratification of this element in the stellar atmosphere. 

As has been pointed out above, for atomic number $Z>20$, all elements are overabundant with the single exception of Ni, which has generally been found deficient in HgMn stars. All iron-peak elements -- Ti, V, Cr, Mn, Fe, Cu, and Co -- are enhanced in HD\,66051. Ti and Cu have been regularly found overabundant in HgMn stars. Interestingly, the observed abundance of Mn is lower than those of Cr and Fe; the odd-$Z$ anomalous abundance pattern observed in the Cr-Mn-Fe triplet and typically encountered in HgMn stars is therefore not seen in HD\,66051. In this respect, it is noteworthy that Y is of lower abundance than Sr and Zr, so no odd-$Z$ anomaly has been observed for the Sr-Y-Zr triplet, too. These finds are reminiscent of the highly peculiar star HD\,65949 \citep{cowley10}. Ga, which is usually significantly enhanced in HgMn stars, is also overabundant in HD\,66051. We note significant overabundances of Au and, in particular, Pt. The Au abundance, however, was estimated from one very weak line ($4016.07$\,{\AA}). 

Lines of the noble gases Ar and Xe are significantly enhanced. Xe, in particular, is strongly enhanced ($\log{\rm\,Xe/Ntot}$ $= -5.07$) -- much more so than its neighbouring element Ba, whose abundance is only slightly higher than solar. The overabundance of Xe and the observed discrepancy between the abundances of Xe and Ba are characteristic of HgMn stars. There are no standard neutron capture r- and s-processes which produce such severe fractionation. The overabundance of Ne has been determined from eight lines and is not significant, which is in agreement with results from the literature that indicate Ne abundances close to the solar value for HgMn stars. 

All analysed lanthanide lines are strongly enhanced in HD\,66051, with the exception of Tm that shows no lines in the investigated spectrum. No atomic data for Pm exists in the public databases, so this element was not considered. In contrast to HgMn stars (see e.g. the considered HD\,49606 and HD\,175640), for which enhancements of Au, Pt, and Hg are rather usual, almost all the REE elements are present and significantly overabundant in the spectrum of HD\,66051. In this respect, the star is very similar to HD\,65949.

\subsection*{Secondary component}

We have identified numerous lines in the composite HARPS spectrum which are not observed in the HIDES spectrum of the primary star, namely various lines of \ion{Fe}{i}, the strongest lines of \ion{C}{i}, \ion{Mg}{i}, \ion{Al}{i}, \ion{Ca}{i}, \ion{Mn}{i}, \ion{Cr}{i}, as well as lines of \ion{Sc}{ii}, \ion{Ti}{ii}, \ion{Cr}{ii}, \ion{Fe}{ii}, \ion{Sr}{ii} and \ion{Y}{ii}. These lines -- some of which are single, unblended lines -- appear redshifted by 55\,\kms\ (+0.825\,\AA\ at 4500\,\AA). We attribute them to the secondary component.

\begin{figure}[ht]
\includegraphics[width=0.90\textwidth]{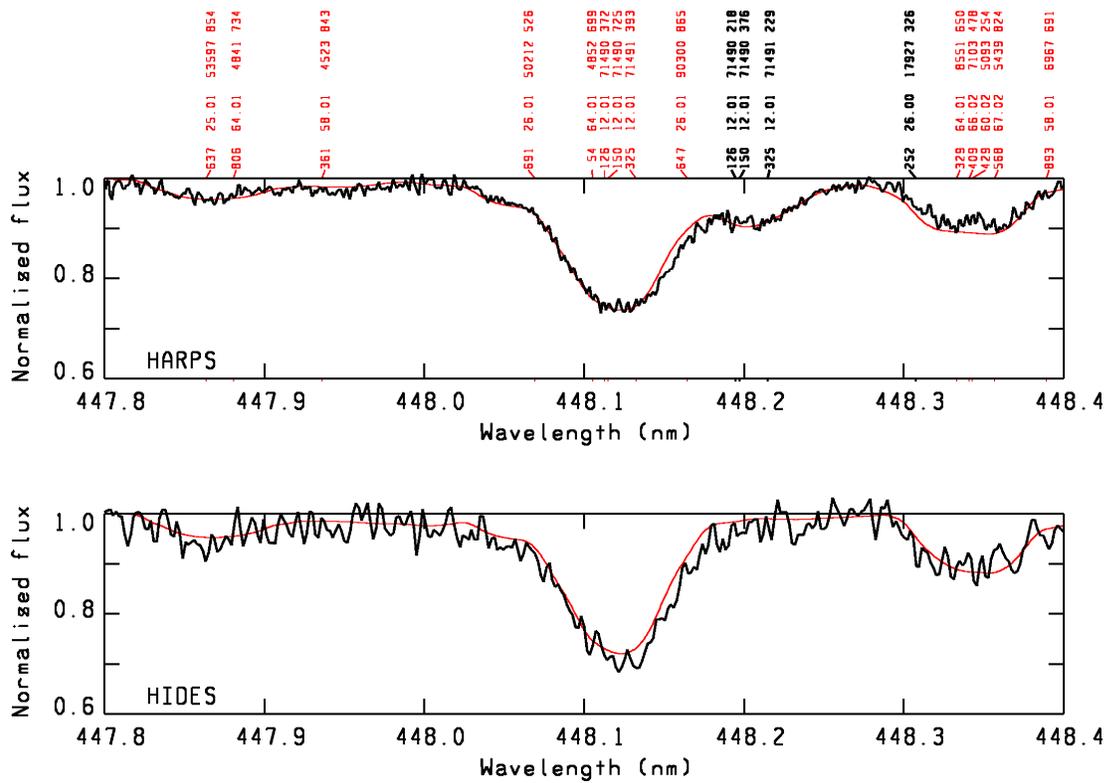}
\caption{The \ion{Mg}{ii} region from 447.8\,nm to 448.4\,nm of the HARPS and HIDES spectra, indicating some of the lines identified in the primary star (red identifications) and the secondary component (black identifications). The observed spectra are shown in black, the synthetic spectra are indicated by the red line.}
\label{fig:synthetic_spectrum}
\end{figure}

On the basis of the observed line spectrum, a model atmosphere with \teff\,=\,$8000$\,K, \logg\,=\,$4.0$, microturbulence \micro\,=\,$2$\,\kms\ and solar abundances was assumed for the secondary star. This model and a projected rotational velocity of \vsini\,=\,$18$\,\kms\ give rise to a synthetic spectrum that, after convolution with the synthetic spectrum of the primary star, satisfactorily reproduces the HARPS spectrum. As an example, Fig.\,\ref{fig:synthetic_spectrum} illustrates the \ion{Mg}{ii} region (447.8\,nm to 448.4\,nm) of the HARPS and HIDES spectra, with line identifications for both components.

At this point, with the available spectroscopic material, it is not possible to make more precise statements about the nature of the secondary component. This will be the topic of an upcoming investigation.

\section*{Discussion}

The chemical composition of HD\,66051 is unique. While the observed abundance pattern follows the general characteristics of the HgMn stars, which are known to exhibit highly individualistic spectra and huge discrepancies in the abundances for a given element from one star to the next, some marked peculiarities are present, most notably the strong silicon overabundance and the lack of the odd-$Z$ anomalies in the Cr-Mn-Fe and Sr-Y-Zr triplets. This pattern is reminiscent of the highly peculiar star HD\,65949 \citep{cowley10}, which also shows a similar overall abundance pattern to our target star. While for HD\,65949, final abundances were determined for 41 elements \citep{cowley10}, we are able to provide abundance values for $46$ elements in HD\,66051.

\citep{cowley10} have put forth the hypothesis that the composition of HD\,65949 might have been influenced by accretion of exotic r-processed material (i.e. material synthesized by the so-called r-process occurring in core-collapse supernovae) that was subsequently subjected to differentiation by atomic diffusion processes. The idea that mass transfer might (also) be involved in the formation of CP star (in particular HgMn star) anomalies has been recurring in the literature for a long time \citep[see][]{wahlgren95}. \citep{cowley10} have developed some interesting assumptions and predictions to be checked against further observations. Interestingly, the anomalies relevant to this hypothesis are also present in HD\,66051. \citep{cowley10} proposed that HD\,65949 might be in a rare intermediate evolutionary state: certain peculiarities common to HgMn stars have not had the time to establish themselves via atomic diffusion. For instance, the N deficiency and the overabundance of P have been proposed to develop before the appearance of a significant Mn excess. Interestingly, Mn is rather low in HD\,66051, too. Another point in case is the shared absence of the high Ga abundance typical of many HgMn stars. According to the speculations of the aforementioned investigators, the Ga anomaly might be a pure diffusion anomaly that had not have the time to develop in HD\,65949. Following these suggestions, HD\,66051 might be in a similar (evolutionary) state as HD\,65949.

Of course, as \citep{cowley10} stress, atomic diffusion may found to be entirely sufficient to explain the abundance pattern of HD\,65949 once the necessary atomic data are known. However, whatever the source of the observed anomalies in both stars, a similar pattern is present. HD\,66051 is therefore well suited to investigate the aforementioned authors' assumptions in more detail and a different object.

By a fortunate twist of fate, and in contrast to HD\,65949, HD\,66051 is an eclipsing binary system. Binaries allow the derivation of fundamental stellar parameters like mass and radius, which is very valuable for CP stars, because very few of them have direct determinations of these parameters \citep{north04}. HD\,66051 therefore lends itself well to follow-on studies concerned with the solution of the system and determination of exact stellar parameters for both components.

The observed binarity opens up several other intriguing possibilities, which are outlined in the following:

\begin{itemize}

	\item Doppler imaging allows the study of the abundance structures and inhomogeneities of CP stars in detail. It enables the mapping of stellar surface inhomogeneities via conversion of line profile time series and offers valuable insights in the physical processes involved \citep[for a short overview, see][]{kochukhov17}. However, some ambiguities are inherent to this technique \citep[e.g.][]{stift17}. These could be resolved, and the Doppler imaging results complemented, by 'eclipse mapping', which allows mapping the surface distribution of abundances by high time-resolution spectroscopy around the time of eclipses in a binary system \citep{north04}. The system of HD\,66051 provides a unique opportunity in this respect.

	\item Open questions concerning the mechanisms at work in producing the abundance anomalies seen in HgMn stars remain \citep{schoeller10}. Atomic diffusion processes are intertwined with parameters like rotation and binarity. Given the high fraction of binaries among HgMn stars, binarity, in particular, has been proposed to play a key role in the understanding of the observed abundance patterns. HD\,66051 may provide a valuable testbed for the further development of these theories.

	\item Some elements in HgMn stars (e.g. Sr, Y, Pt, Zr, and Hg) are suspected to be concentrated in high-altitude clouds. Using high-resolution spectroscopy around the times of eclipses, one might probe different atmospheric layers and, perhaps, directly investigate this possibility.

	\item Generally applicable insight on early-type star formation in binary systems can be derived from studying multiple systems containing peculiar stars \citep{abt73}. According to the aforementioned authors, close-binary formation should be inhibited by the presence of strong magnetic fields. Furthermore, magnetic fields should be dissipated in close systems. However, as discussed below, there are good arguments to assume the presence of a magnetic star in the (moderately close) system of HD\,66051.

\end{itemize}

The most intriguing feature in the light curve of HD\,66051 is the obvious presence of long-lived, secondary variability with an amplitude of $\sim$0.04\,mag in all filters, which we interpret as being due to rotational modulation by a synchronously-rotating star with an inhomogeneous surface structure. However, at this point, the physical mechanism responsible for the observed variability, as well as the system component to which it is associated, remain unclear.

Rotational variability due to cool star-spots is a ubiquitous feature of lower main-sequence stars with convective envelopes below their photospheres \citep{wilson78}. This phenomenon, however, is not to be expected in an early-type main sequence star and -- even less so -- in a CP star. Neither component, therefore, seems likely to exhibit this kind of star-spot-induced variability. It has to be noted, though, that -- using ultra-precise space photometry -- Balona and co-workers have recently put forth evidence for activity in early-type main sequence stars and suggested they may show star-spots in the same way as their cooler counterparts \citep{balona13}. This is still a matter of controversy, though, and the observed amplitudes are very small, in contrast to what has been observed for HD\,66051. Furthermore, as star-spots form and decay, their configuration is expected to change, which is in contrast to the observed stability of the secondary variability in our programme star. The above invoked scenario seems therefore unsuited to explain the observed secondary variability, regardless of the component it is associated to.

HgMn stars have been shown to possess chemical surface inhomogeneities but the occurrence of rotational variability in these stars is a matter of debate \citep{morel14}. In fact, HgMn stars are only rarely found to be photometric variables; the few known cases exhibit very small amplitudes and have been mostly identified using space-based photometry. Furthermore, the underlying causes of the observed variability in these objects have not yet been established. If the observed variability in the light curve of HD\,66051 is to be attributed to the primary star, it would be the first HgMn-related object showing high-amplitude rotational variability. 

On the other hand, the observed, very stable rotational variability fits the light changes expected in the magnetic Ap stars, which are due to chemical abundance inhomogeneities ('abundance spots'). The most natural explanation, then, would be the assumption that the secondary component is actually an Ap star. So far, HD\,161701 is the only known case of a binary system formed by an Ap and an HgMn star \citep{gonzalez14,hubrig14}. However, while the secondary star is situated in the temperature regime of Ap stars, no hint of Ap-star-typical peculiarities are present in our spectroscopic material. High-resolution spectroscopy around the time of the primary minimum (when the secondary component passes in front of the primary star) will be able to shed light on this matter. However, assuming the secondary component contributes only about 40\,\% to the system's optical flux output, we would expect a total amplitude of $>$\,0.1\,mag for the variability of the secondary star in order to produce the observed changes in the (combined) light curve. This would be rather high for an Ap star, as these objects generally exhibit variability with amplitudes of several hundredths of a magnitude. However, cases reaching amplitudes of about 0.1\,mag are known \citep[e.g.][]{bernhard15}.

It is interesting to point out that the primary star exhibits enhanced Si lines, which is unusual for an HgMn star. On the other hand, overabundances of Si are commonly observed in Ap stars and it is well known that Si spots are optically active and at the root of the light variability in many of these stars \citep{krticka07}. It is intriguing to surmise that, perhaps, Si spots on the surface of the primary component might be responsible for the observed rotational variability. The formation of such spots, however, is generally associated with the presence of globally-organized and strong ($>$300\,G and more) magnetic fields, which are not known for HgMn stars. Phase-resolved spectroscopy and polarization measurements of both components would be highly desirable to search for the presence of magnetic fields and establish the true nature of the observed secondary variability in HD\,66051.

To sum up, HD\,66051 offers the unique combination of a highly peculiar HgMn-related star in an eclipsing binary system that shows obvious, rotationally-induced secondary variability. As has been expanded on above, this rare configuration opens up many intriguing possibilities for future research, which might eventually and significantly contribute to the understanding of such diverse phenomena as atmospheric structure, mass transfer, magnetic fields, photometric variability and the origin of the chemical anomalies observed in HgMn stars and related objects.

The analysis of peculiar or outstanding objects often leads to generally applicable insights that help towards an understanding of the broader picture. (24) expressed the hope that the chemically-related star HD\,65949 may prove to be a keystone in understanding the origins of the peculiar composition of the HgMn stars and related objects. We hope that the unique configuration of HD\,66051 might eventually and significantly contribute to this understanding.


\section*{Methods} \label{methods} 

\subsection*{Photometric data -- acquisition and analysis}
The photometric data used in this paper were acquired at the Remote Observatory Atacama Desert \citep[ROAD;][]{hambsch12} with an Orion Optics, UK Optimized Dall Kirkham 406/6.8 telescope and a FLI 16803 CCD camera. Data were obtained through Astrodon Photometric $BVI_{C}$ filters. With 8\,s exposure time in the $3 \times 3$ binning mode, a total of $7850$ measurements were acquired. Twilight sky-flat images were used for flatfield corrections. The reductions were performed with the MAXIM DL program \citep{MAXIM} and the determination of magnitudes using the LesvePhotometry program \citep{LESVE}.

The photometric data were cleaned of obvious outliers by visual inspection and analysed using the software package \textsc{Period04} \citep{period04} which is based on a discrete Fourier transform algorithm. The orbital period solution was improved on manually.

\subsection*{Spectroscopic analysis}

The HARPS spectrum of HD\,66051 ($\rm{R}\sim110\,000$, spectral range $390$\,nm to $690$\,nm) was procured from the archive of the 'Variable Star One-shot Project'. Details on the spectroscopic observations can be found in the corresponding publication \citep{dall07}. The HIDES spectrum was taken to decide which lines of the HARPS spectrum belong to the primary star and which can be used to analyse the secondary star. It covers the spectral range from $409$\,nm to $752$\,nm with the resolving power $\rm{R}\sim50\,000$. The reduction was made using dedicated IRAF-based scripts \citep{2016MNRAS.461.2896H}. Both spectra have been normalized to the continuum using the standard {\sc iraf} procedure {\it continuum}.

In the spectroscopic classification, Si peculiarity was established using the lines at $4128$\,{\AA}, $4131$\,{\AA} (blended with Eu), $4028$\,{\AA}, and $4076$\,{\AA} (blended with Sr and Cr). Cr peculiarity was assumed from the enhanced line at $4172$\,{\AA} and established using the $4111$\,{\AA} line. Enhancement of Sr was recognized at $4077$\,{\AA} and confirmed via the $4216$\,{\AA} line. Eu enhancement was investigated using the lines at $4130$\,{\AA} and $4205$\,{\AA}, which are strongly blended with the lines of other elements; therefore, care has to be taken in the interpretation of this result.

The necessary atmospheric models (plane-parallel, hydrostatic and radiative equilibrium, 1-dimensional) were calculated with the {\small\sc ATLAS\,9} code \citep{KuruczR}, ported to GNU/Linux by \citep{sbordone05}. The grid of atmospheric models was calculated for effective temperatures from $11000$ to $14000$\,K with a step of $100$\,K, surface gravities from $3.0$ to $4.3$\,dex with a step of $0.1$\,dex, microturbulence velocities between $0.0$, $0.5$ and $1.0$\,km\,s$^{-1}$, and for metallicities [M/H] equal to $0.0$, $0.5$ and $1.0$\,dex. The appropriate synthetic spectra were computed with the Fortran {\small\sc SYNTHE} code \citep{KuruczR}, which calculates intensity stellar spectra for a given model atmosphere under the assumption of LTE.

The atomic data available on the Castelli website \citep{FIORELLA} were supplemented for the second and third spectra of the lanthanides with the data taken from the Vienna Atomic Line Database \citep[VALD,][]{kupka99}, originally presented in the Data on Rare Earths At Mons University (DREAM) database \citep{biemont99}.

\subsection*{Atmospheric parameters}

Effective temperature (\teff) and surface gravity (\logg) were calculated from calibrations of photometric indices and have been treated as the initial parameters for the spectroscopic analysis, during which a synthesis of hydrogen line profiles has been used to obtain these parameters. The methods employed are described in the following.

\begin{itemize}
\item \textbf{Str\"{o}mgren {\it uvby$\beta$} photometry}: The $uvby\beta$ indices were taken from the catalogue of \citep{paunzen15}. Atmospheric parameters were determined using the {\small\sc UVBYBETA} code written by \citep{moon85} and corrected by \citep{napiwotzki93}. From the $c_1$ and $\beta$ indices, we obtained an effective temperature of \teff\,=\,$13200$\,K and surface gravity \logg\,=\,$4.40$\,dex, while the $\rm{[u-b]}$ index is indicative of \teff\,=\,$12240$\,K. The interstellar reddening $E(b-y)$ was determined as $0.018$\,mag. However, because of the increased line-blanketing in CP stars, these values are only approximate solutions and further corrections need to be taken into account. We therefore applied the corrections for helium-weak CP stars given by \citep{netopil08}, which resulted in an effective temperature of \teff\,=\,$12910$\,K as derived from the $c_1$ and $\beta$ indices. The somewhat lower temperature of \teff\,=\,$12050$\,K was calculated from the $\rm{[u-b]}$ index according to the relation \teff\,=\,$5040/(0.173+0.286*[u-b])$ given by \citep{netopil08}. We have adopted errors of $\pm300$\,K for both temperature values, as advised by the aforementioned author.

\item \textbf{Geneva photometry}: Geneva photometry of our programme star was retrieved from the General Catalogue of Photometric Data database \citep{hauck98}. To determine atmospheric parameters, we have employed the codes and calibrations performed by \citep{kunzli97}, which can be applied to B, A, and F stars with luminosity classes V-III and are based on LTE Kurucz atmosphere models. To calculate atmospheric parameters from Geneva photometry, knowledge of the interstellar reddening is necessary. The colour excess $E(B-V)$ was determined from $E(b-y) = 0.018$\,mag by employing the relation $E(B-V) = 1.43*E(b-y)=0.026$\,mag \citep{netopil08}. Values have been calculated assuming different metallicities. For solar metallicity ({\rm [M/H]}\,=\,0.0\,dex), we have determined \teff\,=\,$13010\pm70$\,K and \logg\,=\,$4.20\pm0.10$. For \rm{[M/H]}\,=\,1.0\,dex, we have estimated \teff\,=\,$12750\pm60$\,K and \logg\,=\,$4.25\pm0.10$\,dex. Taking into account the corrections given by \citep{netopil08}, lower effective temperatures of \teff\,=\,$12725\pm285$\,K and \teff\,=\,$12490\pm255$\,K have been calculated for {\rm [M/H]}\,=\,0.0\,dex and \rm{[M/H]}\,=\,1.0\,dex, respectively.

\item \textbf{Hydrogen Balmer line fitting}: Using the sensitivity of the Balmer H$\alpha$ and H$\beta$ lines to effective temperature and surface gravity, we have obtained \teff\,=\,$12500\pm200$\,K and \logg\,=\,$4.0\pm0.1$\,dex. We have selected these two hydrogen lines for the analysis because they are less contaminated by the metal lines and are easier to normalize. To derive atmospheric parameters, we used an iterative approach which minimizes the differences between observed and synthetic H$\alpha$ and H$\beta$ profiles \citep[see][]{catanzaro04}. To estimate the error of \teff\ and \logg\, we took into account the differences in the obtained \teff\ and \logg\ values from separate Balmer lines, resulting from validity of normalization.

\end{itemize}

\section*{Data availability}
The multicolour photometric observations acquired in this study are available as supplementary material to this paper. The HARPS spectrum is available in the ESO Telescopy Bibliography repository  (http://telbib.eso.org/detail.\-php?bibcode=2007A\%26A...470.1201D). The analysed HIDES spectrum is available as supplementary material to this paper.


\bibliography{HD66051_SciRep_final}


\section*{Acknowledgements}
EN and KH acknowledge the Polish National Science Centre grants no. 2014/13/B/ST9/00902 and 2016/21/B/ST9/01613. Calculations have been carried out at the Wroc{\l}aw Centre for Networking and Supercomputing (http://www.wcss.pl), grant No.\,214. Atomic data compiled in the DREAM database (E. Biemont, P. Palmeri \& P. Quinet, Astrophys. Space Sci. 269-270, 635, 1999) were extracted via VALD (Kupka et al., 1999, A\&AS 138, 119, and references therein).

\section*{Author contributions}
E.N. performed the spectral classification, determined atmospheric parameters and chemical abundances of the primary component of the system, described the results, contributed to the manuscript text, and prepared figures 1, 3, 4, and 5;
S.H. worked on the interpretation of the data and wrote the main manuscript text;
F.C. checked the atmospheric parameters and chemical abundances of the primary component, performed spectroscopic analysis of the secondary component and prepared figure 2;
S.H., E.P. and K.B. initially identified the peculiar nature of HD\,66051, performed a preliminary spectral classification, helped with the interpretation of the data and set up collaboration;
F.-J.H. acquired photometric observations of the star in $BVI_C$ bands with the remote observatory ROAD and reduced the data;
K.H. made spectroscopic observations of the star with the HIDES instrument and reduced and calibrated the data.
All authors reviewed the manuscript.

\section*{Competing financial interests}
The authors declare that they have no competing financial interests.

\begin{table}[ht]
\begin{tabular}{|l|l|l|}
\toprule
Method                               & Result                                              & Remark                                   \\ \midrule
Str{\"o}mgren $uvby\beta$ photometry & \teff\,=\,$12910\pm300$\,K; \logg\,=\,$4.40$        & from $c_1$ and $\beta$ indices           \\
                                     & \teff\,=\,$12050\pm300$\,K                          & from $\rm{[u-b]}$ index                  \\ \midrule
Geneva photometry                    & \teff\,=\,$12725\pm285$\,K; \logg\,=\,$4.20\pm0.10$ & for \rm{[M/H]}\,=\,0.0\,dex              \\
                                     & \teff\,=\,$12490\pm255$\,K; \logg\,=\,$4.25\pm0.10$ & for \rm{[M/H]}\,=\,1.0\,dex              \\ \midrule
Hydrogen line profile fitting        & \teff\,=\,$12500\pm200$\,K; \logg\,=\,$4.00\pm0.10$ & from H$\alpha$ and H$\beta$ lines        \\ \midrule \midrule
\textit{(Spectral synthesis)}		 & \textit{\teff\,$\sim$\,8000\,K; \logg\,=\,4.0; }    & \textit{estimate for the secondary star} \\
														    & \textit{\vsini\,=\,18\,\kms} &                                          \\
\bottomrule
\end{tabular}
\caption{Atmospheric parameters of HD\,66051, as derived from different methods. The first three rows list parameters of the primary component, the bottom row (highlighted by italic font) indicates an estimate for the secondary star.}
\label{atmospheric_parameters}
\end{table}

\begin{table}[ht]
\begin{tabular}{|l|l|c||l|l|c|}
\toprule
Element    & Abundance          & Solar     & Element    & Abundance                & Solar     \\
           &                    & abundance &            &                          & abundance \\
\midrule

He (2)     &$-2.50\pm0.10$ (5)  &   $-1.11$ &       Zr (40)    &$-7.10\pm0.25$ (9)  &  $ -9.46$ \\           
C  (6)     &$-4.52\pm0.10$ (4)  &   $-3.61$ &       Nb (41)    &$-7.22\pm0.30$ (3)  &  $-10.58$ \\            
O  (8)     &$-3.71\pm0.20$ (5)  &   $-3.35$ &       Mo (42)    &$-7.12\pm0.20$ (4)  &  $-10.16$ \\           
Ne (10)    &$-3.80\pm0.30$ (8)  &   $-4.11$ &       Ru (44)    &$-7.80       $ (1)  &  $-10.29$ \\
Mg (12)    &$-5.36\pm0.20$ (4)  &   $-4.44$ &       Xe (54)    &$-5.07\pm0.25$ (3)  &  $ -9.80$ \\
Al (13)    &$-7.07\pm0.40$ (6)  &   $-5.59$ &       Ba (56)    &$-9.71       $ (1)  &  $ -9.86$ \\
Si (14)    &$-3.55\pm0.22$ (28) &   $-4.53$ &       Ce (58)    &$-7.00\pm0.40$ (6)  &  $-10.46$ \\
P  (15)    &$-5.40\pm0.20$ (8)  &   $-6.63$ &       Pr (59)    &$-6.44\pm0.28$ (20) &  $-11.32$ \\
S  (16)    &$-4.76\pm0.39$ (7)  &   $-4.92$ &       Nd (60)    &$-6.02\pm0.30$ (25) &  $-10.62$ \\
Cl (17)    &$-4.74\pm0.10$ (5)  &   $-6.54$ &       Sm (62)    &$-7.14\pm0.38$ (5)  &  $-11.08$ \\
Ar (18)    &$-5.24\pm0.20$ (5)  &   $-5.64$ &       Eu (63)    &$-7.29\pm0.40$ (3)  &  $-11.52$ \\
Ca (20)    &$-5.25\pm0.20$ (6)  &   $-5.70$ &       Gd (64)    &$-6.75\pm0.18$ (5)  &  $-10.97$ \\
Ti (22)    &$-5.44\pm0.32$ (20) &   $-7.09$ &       Tb (65)    &$-7.63\pm0.20$ (19) &  $-11.74$ \\
V  (23)    &$-7.00\pm0.20$ (3)  &   $-8.11$ &       Dy (66)    &$-6.37\pm0.38$ (14) &  $-10.94$ \\
Cr (24)    &$-5.35\pm0.35$ (26) &   $-6.40$ &       Ho (67)    &$-6.96\pm0.30$ (9)  &  $-11.56$ \\
Mn (25)    &$-5.70\pm0.20$ (16) &   $-6.61$ &       Er (68)    &$-7.18\pm0.32$ (11) &  $-11.12$ \\
Fe (26)    &$-3.50\pm0.20$ (109)&   $-4.54$ &       Yb (70)    &$-7.00\pm0.30$ (4)  &  $-10.20$ \\
Co (27)    &$-5.60\pm0.30$ (3)  &   $-7.05$ &       Lu (71)    &$-7.20\pm0.20$ (3)  &  $-11.94$ \\
Ni (28)    &$-6.20\pm0.20$ (10) &   $-5.82$ &       Hf (72)    &$-7.95\pm0.28$ (12) &  $-11.19$ \\
Cu (29)    &$-5.73\pm0.20$ (4)  &   $-7.85$ &       Pt (78)    &$-5.94\pm0.30$ (3)  &  $-10.42$ \\
Ga (31)    &$-6.50\pm0.30$ (3)  &   $-9.00$ &       Au (79)    &$-7.03       $ (1)  &  $-11.12$ \\
Sr (38)    &$-6.20\pm0.20$ (4)  &   $-9.17$ &       Hg (80)    &$-7.21\pm0.20$ (4)  &  $-10.87$ \\
Y  (39)    &$-7.80\pm0.30$ (4)  &   $-9.83$ &       Pb (82)    &$-6.79       $ (2)  &  $-10.29$ \\

\bottomrule
\end{tabular}
\caption{Chemical abundances ($\log {\rm N/N_{tot}}$) and standard deviations for individual elements. Number of individual lines and blends analysed is given in parentheses behind the abundance values.
Solar abundances were taken from \citep{asplund09}.}
\label{chemical_abundances}
\end{table}

\end{document}